\documentclass[runningheads]{llncs}

\usepackage{graphicx}
\usepackage{cite}
\usepackage[caption=false]{subfig}
\usepackage{float}
\usepackage{todonotes}
\usepackage{placeins}

\usepackage{hyperref}
\usepackage[title]{appendix}

\begin{document}

\title{The demography of the peripatetic researcher: Evidence on highly mobile scholars from the Web of Science}

%
\titlerunning{The demography of the peripatetic researcher}
%
\author{Samin Aref\inst{1}\orcidID{0000-0002-5870-9253} 
 \and \\
Emilio Zagheni\inst{1}\orcidID{0000-0002-7660-8368} 
 \and \\
Jevin West\inst{2}\orcidID{0000-0002-4118-0322}}
\authorrunning{S. Aref et al.}
%
\institute{
Max Planck Institute for Demographic Research, Konrad-Zuse-Str. 1\\
18057 Rostock, Germany\\
\email{\{aref,zagheni\}@demogr.mpg.de}\\
\and
Information School, University of Washington, Seattle, WA 98195, USA
\email{jevinw@uw.edu}
}
\maketitle              
%
{
	\small
The policy debate around researchers' geographic mobility has been moving away from a theorized zero-sum game in which countries can be winners (``brain gain'') or losers (``brain drain''), and toward the concept of ``brain circulation,'' which implies that researchers move in and out of countries and everyone benefits. Quantifying trends in researchers' movements is key to understanding the drivers of the mobility of talent, as well as the implications of these patterns for the global system of science, and for the competitive advantages of individual countries. Existing studies have investigated bilateral flows of researchers. However, in order to understand migration systems, determining the extent to which researchers have worked in more than two countries is essential. This study focuses on the subgroup of highly mobile researchers whom we refer to as ``peripatetic researchers'' or ``super-movers.'' 

More specifically, our aim is to track the international movements of researchers who have published in more than two countries through changes in the main affiliation addresses of researchers in over 62 million publications indexed in the Web of Science database over the 1956-2016 period. Using this approach, we have established a longitudinal dataset on the international movements of highly mobile researchers across all subject categories, and in all disciplines of scholarship. This article contributes to the literature by offering for the first time a snapshot of the key features of highly mobile researchers, including their patterns of migration and return migration by academic age, the relative frequency of their disciplines, and the relative frequency of their countries of origin and destination. Among other findings, the results point to the emergence of a global system that includes the USA and China as two large hubs, and England and Germany as two smaller hubs for highly mobile researchers.
}

The reference to this article should be made as follows:

 {\scshape Aref, S., Zagheni, E., and West, J.}
	\newblock The demography of the peripatetic researcher: Evidence on highly mobile scholars from the Web of Science.
	\newblock {\em Lecture Notes in Computer Science}, (2019), Proceedings of the 11\textsuperscript{th} International Conference on Social Informatics, Doha, Qatar.


\section{Introduction}\label{s:intro}

In the global economy, highly skilled migration \cite{czaika_high-skilled_2018} and the mobility of researchers \cite{arrieta_quantifying_2017} have become central issues for research and policy. The interest in these issues is reflected in numerous studies that have investigated the mobility of researchers across countries \cite{canibano_international_2011,moed_studying_2013,moed_bibliometric_2014,marmolejo-leyva_mobility_2015,robinson-garcia_scientific_2016,sugimoto_scientists_2017,czaika_globalisation_2018,robinson-garcia_many_2019,vaccario2018reproducing}. In this article, we focus on the movements of researchers as a subcategory of mobility processes among highly educated people, which have far-reaching consequences for the exchange of knowledge and the development of new ideas, as well as for the emergence of competitive advantages for the countries involved in the resulting circulation of knowledge. 

Studying migration among the highly skilled (and researchers in particular) at the global level is difficult using classic demographic methods, in part because a world migration survey does not exist \cite{willekens_international_2016}. Recent studies that have examined international mobility among researchers have used bibliometric data as a complementary approach \cite{moed_bibliometric_2014,czaika_globalisation_2018}. This method involves tracking the international movements of researchers through the changes in the affiliation addresses on their publications. By generating the equivalent of a census of publications, the application of this method makes it possible to assess the mobility of scholars. 

The feasibility of this method has been tested in previous studies that estimated migration flows \cite{moed_studying_2013,moed_bibliometric_2014}. As most of the existing studies on this topic limited their focus to specific disciplines \cite{hadiji_computer_nodate,dyachenko_internal_2017}, or to comparisons of a few countries \cite{moed_studying_2013,conchi_scientific_2014,marmolejo-leyva_mobility_2015,robinson-garcia_scientific_2016,dyachenko_internal_2017} or of mobile vs.\ non-mobile scholars \cite{marmolejo-leyva_mobility_2015,sugimoto_scientists_2017}, more in-depth analysis is needed to provide us with a better understanding of international mobility in academia. That is the objective of this study.

The issue of the mobility of researchers is similar to the issue of migration, as a survey of 17,000 researchers showed that among the factors researchers consider when weighing an international move include the opportunity to achieve a better quality of life and a higher salary, and the desire to discover another culture \cite{franzoni_movers_2014,scellato_migrant_2015,franzoni_international_2015}. The idea of using the affiliations of researchers to analyze their mobility patterns can be traced back to Rosenfeld and Jones in 1987 \cite{rosenfeld1987patterns}, who used biographical information from the American Psychology Association directory to study the mobility patterns of psychologists by gender. Another pioneering study \cite{laudel_studying_2003} tested the suitability of bibliometric data in the biomedical field for studying the international mobility of elite researchers, and for investigating phenomena such as brain drain. Several studies have shown that internationally mobile scholars have a substantially greater research impact than non-mobile scholars when measured by citation-based indicators \cite{sugimoto_scientists_2017, robinson-garcia_many_2019}. More comprehensive studies on this topic investigated return migration among scholars using a 10-year worth of bibliometric data for 17 countries \cite{moed_bibliometric_2014}, and combined bibliometric data with other data sources to examine the relationship between return mobility and scientific impact in the context of Europe \cite{canibano2017inquiry}. Other recent studies have investigated the directional flows of researchers using bibliometric data over a 60-year period ending in 2009 \cite{vaccario2018reproducing}, and have used networks to model and analyze geographical career paths over time \cite{vaccario2019mobility}. 

Our focus here is on researchers who have published with main affiliation addresses from at least three distinct countries, according to Web of Science data over the 1956-2016 period. We refer to these researchers as ``peripatetic researchers'' or ``super-movers'' and provide analyses of their common features, mobility paths, and return migration. 

\section{Methods and Materials}\label{s:mm}
A key advantage of using bibliometric data for studying the mobility of researchers is the availability of millions of publications in bibliometric databases, such as Web of Science \cite{robinson-garcia_many_2019}, Scopus \cite{appelt_which_2015}, and Dimensions \cite{thelwall_dimensions:_2018}. Each publication serves as a data point that indicates the addresses of the authors on a certain publication date. These data points provide proxies not only on the geographic locations of researchers, but on their fields of research and the disciplines of the publication venues. We track the international movements of researchers through the changes in the main affiliation addresses listed in more than 62 million publications indexed in the Web of Science database. This is the initial step in establishing a longitudinal dataset on the mobility of researchers across all research fields, and in different disciplines. Observing, consistent with the literature \cite{robinson-garcia_many_2019}, that more than 90\% of the researchers showed no signs of international mobility, we focus on the small fraction of scholars whose main affiliation track indicates that they moved across international borders. In particular, we focus on researchers whose Web of Science publication data show that they have published with main affiliation addresses from at least three distinct countries, which we consider an indication that they made more than one international move. 

With a nod to Aristotle's Peripatetic school, we refer to this small group of highly mobile scholars as ``\textit{peripatetic researchers}.'' The term ``peripatetic'' means ``moving from place to place.'' Derived etymologically from the Greek, peripatetic literally means ``of walking,'' as Aristotle required his students to walk alongside him as he lectured. Using our dataset of ``peripatetic researchers'' (whom we also call ```\textit{super-movers}''), we provide several in-depth statistics related to mobility and return migration, disaggregated by new variables involving age and origin. Similar methodologies have been deployed in the past, but a longitudinal global-level study that includes all highly mobile scholars has not previously been undertaken. Our aim in conducting such a study is to shed some light on the common characteristics of peripatetic researchers and their mobility patterns. We do so by tracking the international paths of these researchers over the 1956-2016 period. While there are a number of methodological challenges associated with studying scientific mobility and collaboration using bibliometric data that should be taken into consideration~\cite{moed_bibliometric_2014, chinchilla-rodriguez_networks_2017, aref_analysing_2018, czaika_globalisation_2018, alburez2019}, we believe that the novel approach we propose and the results we obtain will prove timely and relevant, and will provide a foundation for future research in this field. 

\section{Results}\label{s:results}
In this section, we present the main results of our analysis of Web of Science authorship records from 1956 to 2016. An authorship record is the linkage between a publication and an author of that publication. We extracted all of the authorship records of the super-movers (the individuals whose publications had main affiliation addresses in at least three distinct countries). This extraction resulted in nearly 1.7 million authorship records, which make up the main dataset that we describe and analyze in the following subsections.

An initial look at the most common countries in the dataset shows that the USA, China, Japan, Germany, England, and South Korea are the six countries with the highest number of authorship records. Of the total 1.7 million authorship records of peripatetic researchers in our dataset, almost 68\% refers to one of these six countries which all have more than 100,000 authorship records. 

\subsection{Common characteristics of peripatetic researchers}\label{ss:overall}
\begin{figure}
\centering
	\subfloat[Countries of academic origin]{
		\includegraphics[width=0.48\textwidth]{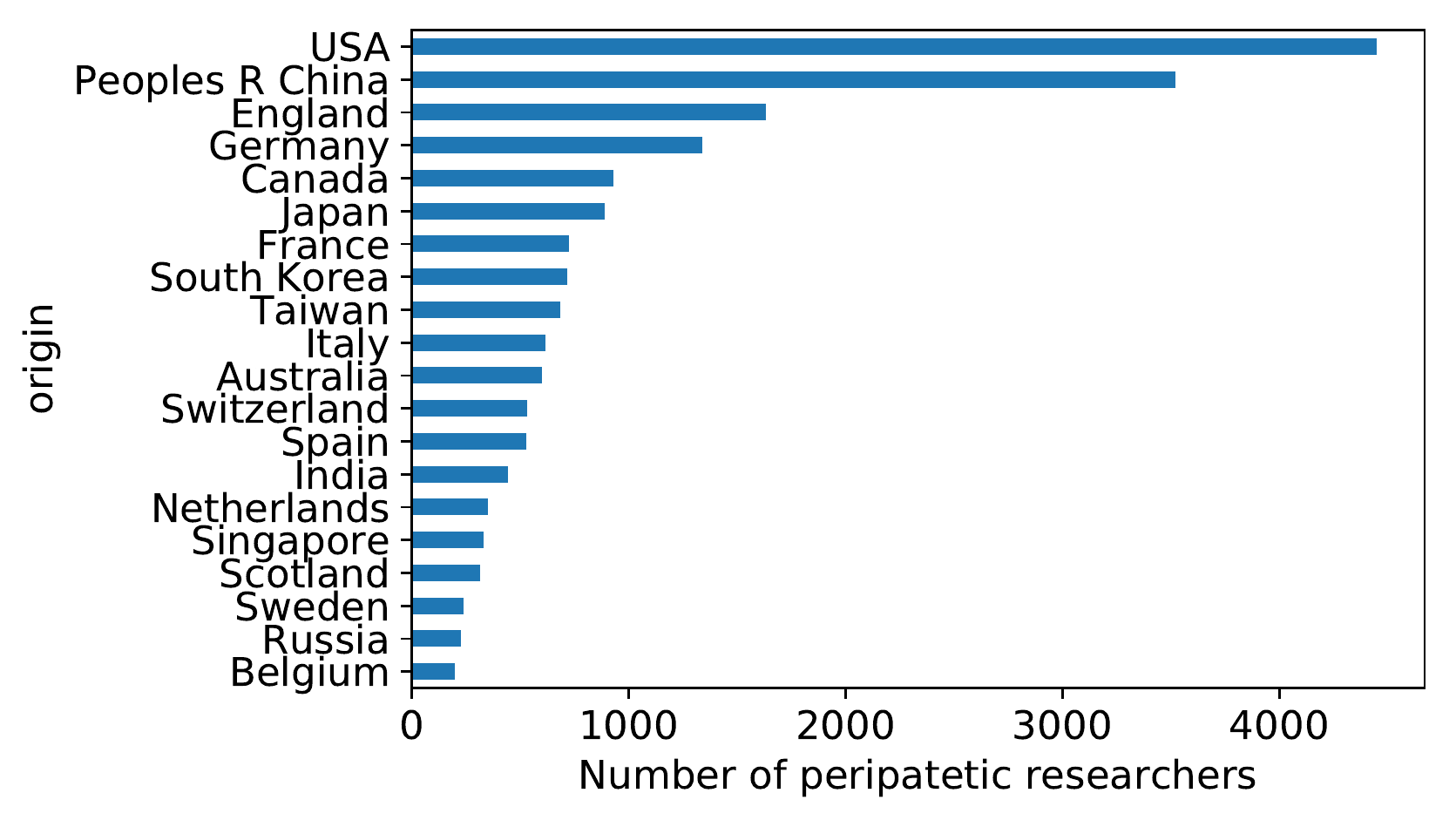}
		\label{subfig:countries_of_origin}
	}
	\subfloat[Countries of academic destination]{
		\includegraphics[width=0.48\textwidth]{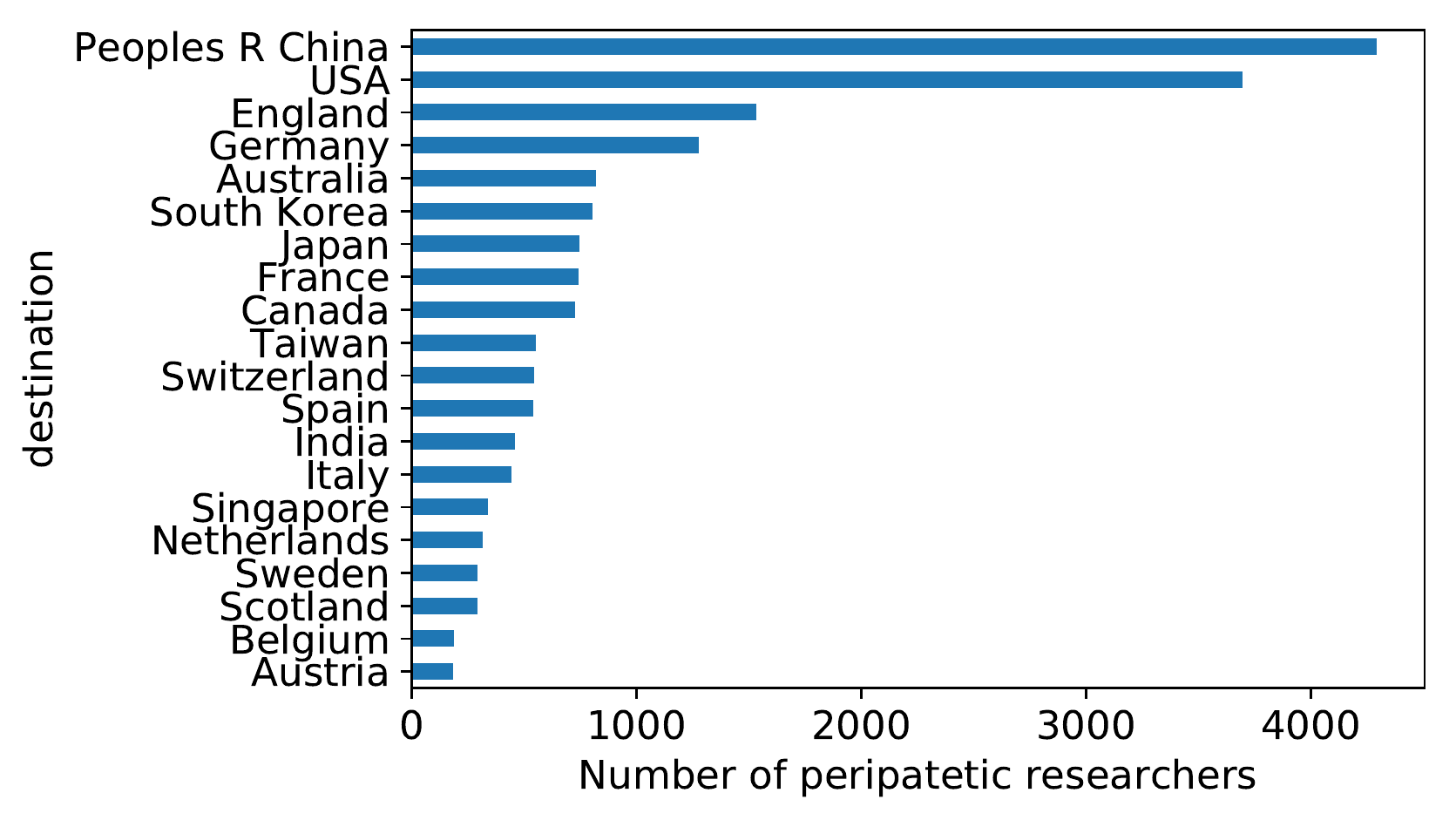}
		\label{subfig:countries_of_destination}
	}
	\caption{Twenty most common countries of academic origin (a) and academic destination (b) among the super-movers} \label{fig:countries_of}
\end{figure}

We define \textit{the country of academic origin (destination)} as the country that appears in the earliest (latest) publication of an individual researcher. Figure ~\ref{fig:countries_of} illustrates the most common countries of main affiliation associated with the earliest publications of the super-movers (in Subfigure ~\ref{subfig:countries_of_origin}), and with their latest publications (in Subfigure ~\ref{subfig:countries_of_destination}) over the 1956-2016 period. Figure~\ref{fig:countries_of} shows that the USA and China were the most common and the second-most common countries of academic origin, while this order was reversed for the destinations of the super-movers. When we take the relationship between mobility and scientific impact into account \cite{sugimoto_scientists_2017, robinson-garcia_many_2019}, we find that our observations for the USA and China are consistent with a \textit{Nature} survey of 2,300 respondents \cite{van_noorden_global_2012}, which showed that although the USA has historically been the country with the greatest scientific impact, China is expected to have the greatest impact in 2020.

Also in Figure~\ref{fig:countries_of}, we can see that England and Germany were, respectively, the third- and fourth-most common countries of origin and countries of destination. Countries such as Canada and Italy were more likely to be the country of academic origin (ranked several steps higher as the country of origin) than the country of destination, while the opposite pattern (ranked several steps higher as the country of destination) is observed for South Korea and Australia. 

We divide the number of super-movers for each country of academic origin by the country's population (in 2016 \cite{UN_WPP2019}) to obtain a measure of super-movers per capita. This allows us to make a different comparison across countries. Of the 20 countries included in Subfigure~\ref{subfig:countries_of_origin}, Switzerland, Singapore, Scotland, England, Taiwan, Canada, and Australia were ranked 1, 2, 3, 6, 7, 9, and 10, respectively, in terms of the number of super-movers per capita. It should be noted that neither the USA nor China were in the top 20 countries based on the per capita measurement. 

\begin{figure}[ht]
\centering
	\subfloat[Subjects of publication venues]{
		\includegraphics[height=0.34\textwidth]{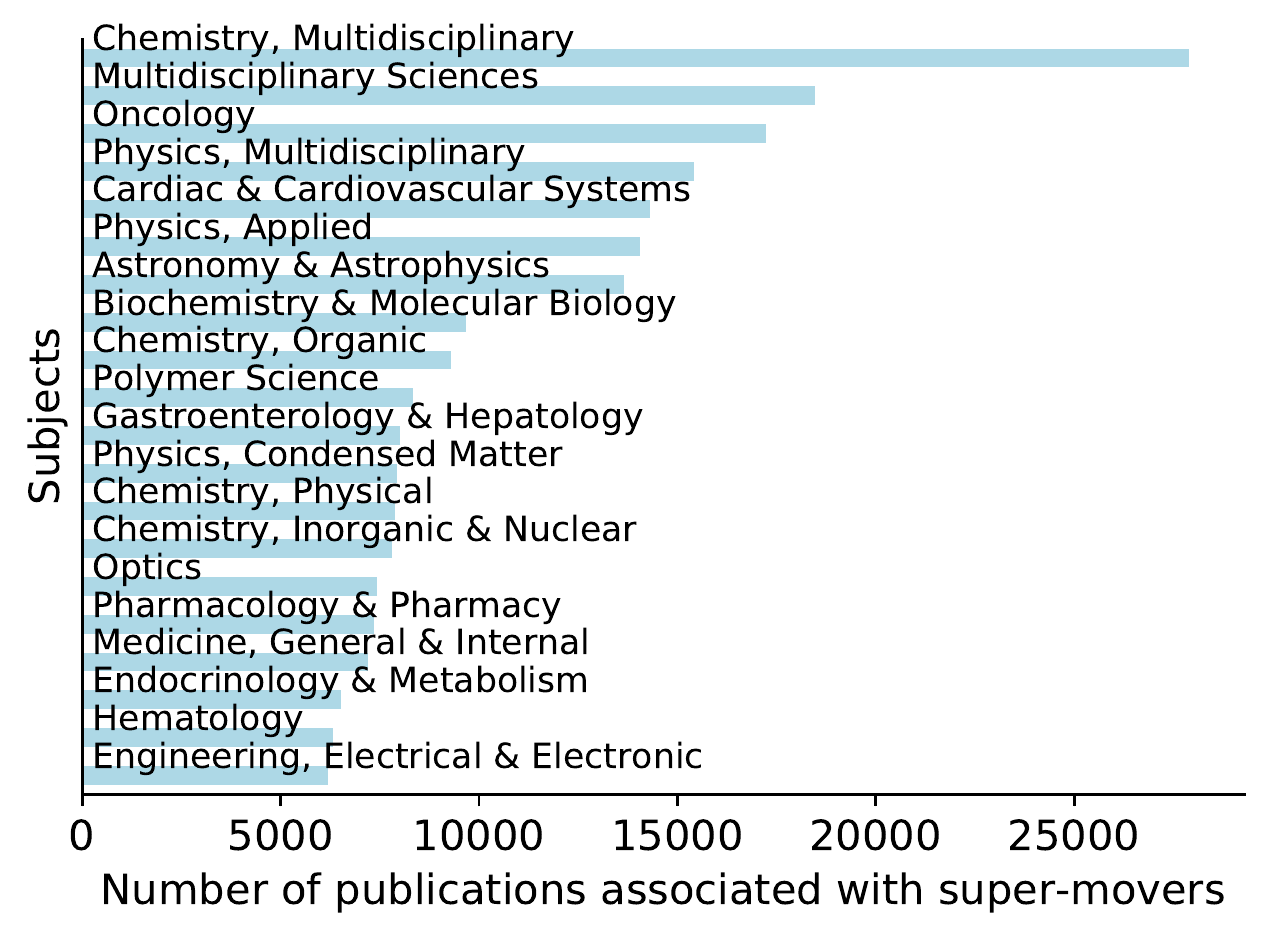}
		\label{subfig:subjects}
	}
	\subfloat[Titles of publication venues]{
		\includegraphics[height=0.34\textwidth]{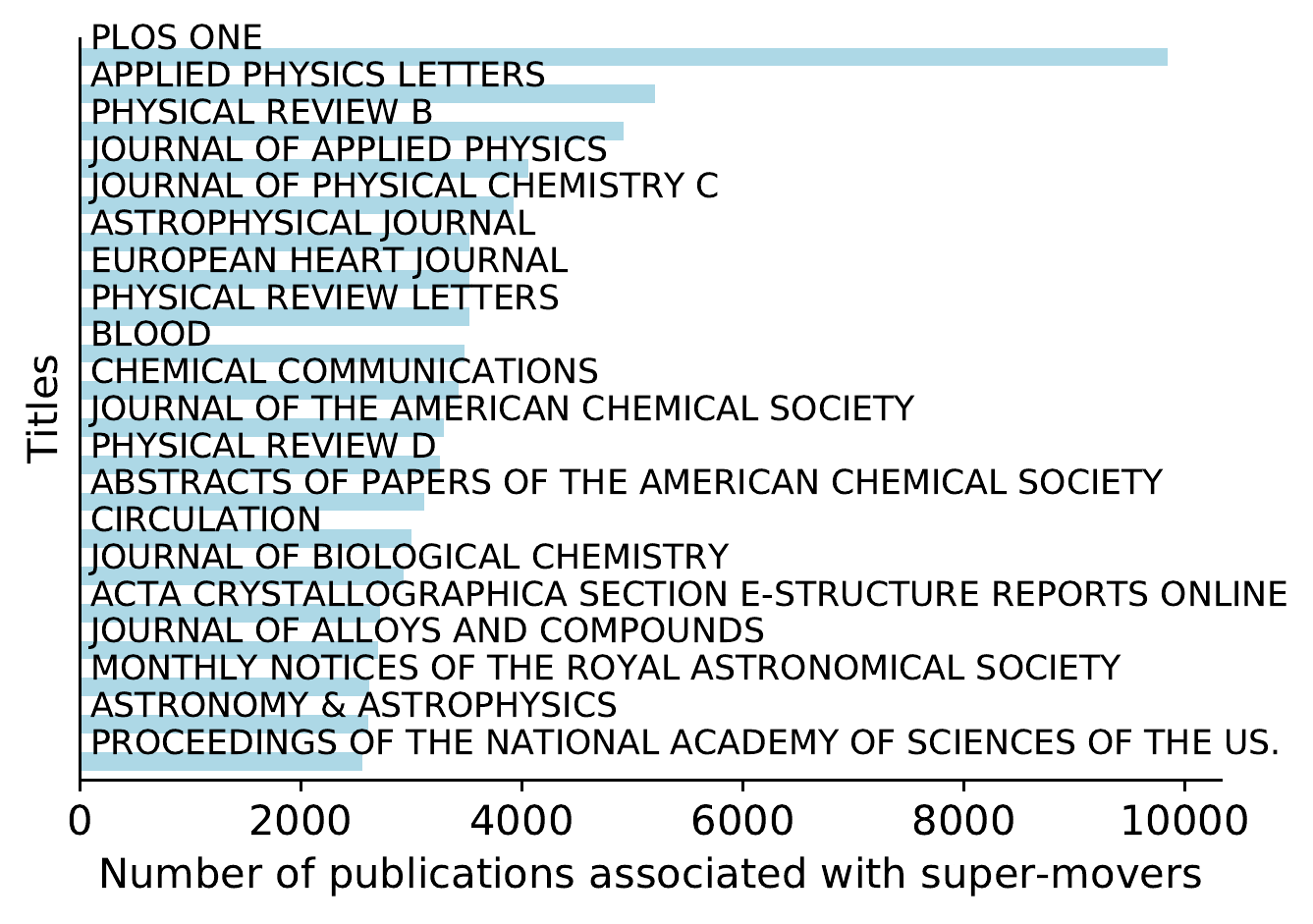}
		\label{subfig:journals}
	}
	\caption{Twenty most common subject categories (a) and titles (b) of publication venues among the publications of the super-movers}
	\label{fig:sources}
\end{figure}

We also investigate the question of whether the scientific output of the peripatetic researchers was homogeneous across disciplines using the titles and subject categories of the publication venues in our dataset of 1.7 million authorship records. Figure~\ref{fig:sources} shows the subject categories and titles of the publication venues that appeared most frequently. Looking at Subfigure~\ref{subfig:subjects}, we can see that multidisciplinary chemistry was by far the most common subject among the authorship records of the super-movers. Among the other subjects that appeared most frequently were multidisciplinary sciences, oncology, and multidisciplinary physics, followed by several other fields of chemistry, physics, and medicine that were ranked in the top 10. An analysis of the most common publication venue titles show a fairly similar pattern, with \textit{Plos One} being ranked first followed by several physics journals, alongside \textit{European Heart Journal}, \textit{Blood}, and \textit{Chemical Communications} which were ranked in the top ten. Meanwhile, several chemistry and physics journals, \textit{Circulation}, and \textit{Journal of Alloys and Compounds} appeared in the lower ranks of the top 20 list displayed in Subfigure~\ref{subfig:journals}. 

\begin{figure}[hbt!]
	\centering
		\includegraphics[height=0.36\textheight]{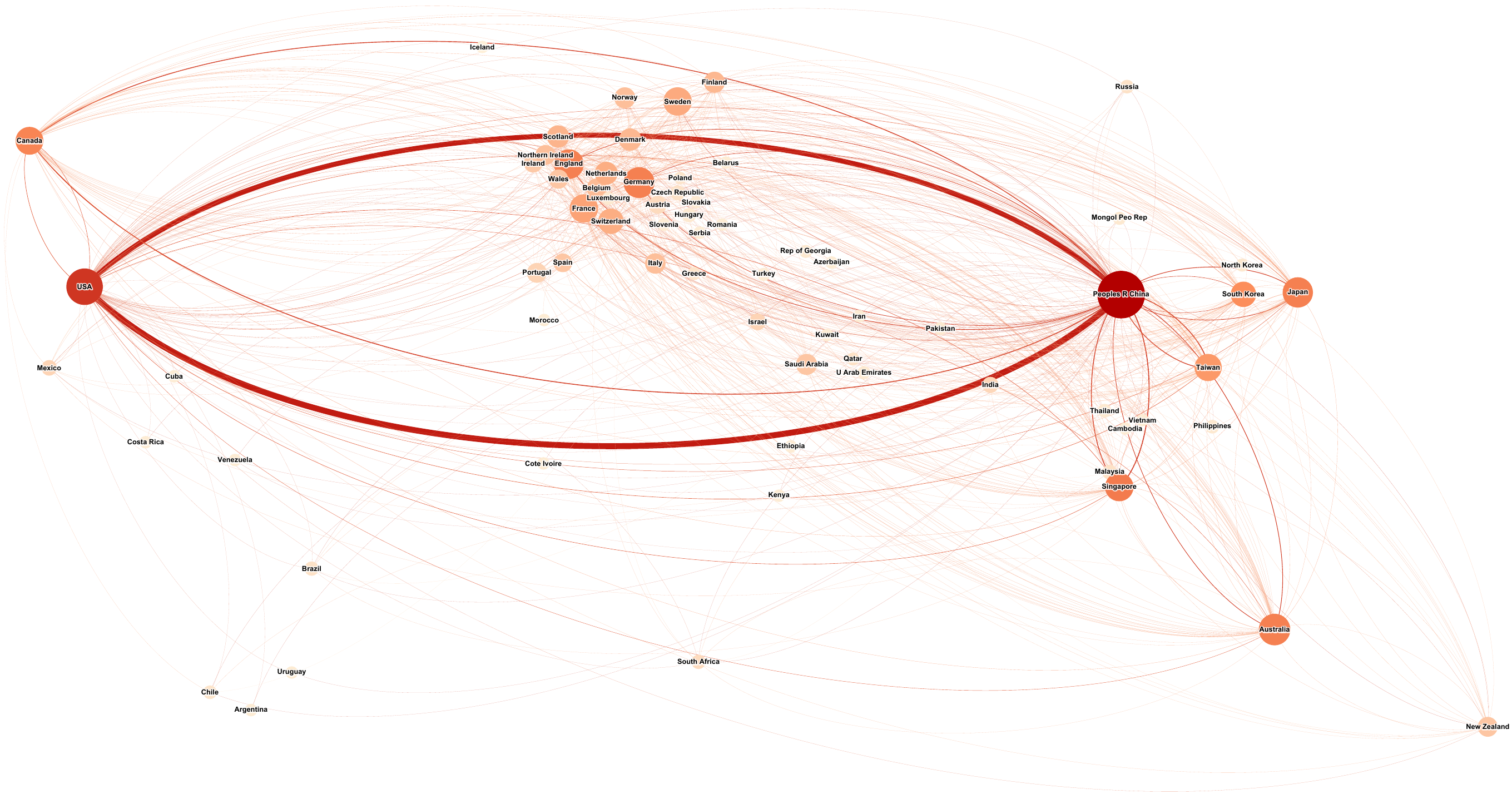}
		\label{fig:china}
		\caption{Paths for super-movers from country of academic origin China (high-resolution version online)}
\end{figure}

\begin{figure}
\centering
		\includegraphics[angle=90,origin=c,width=1\textwidth]{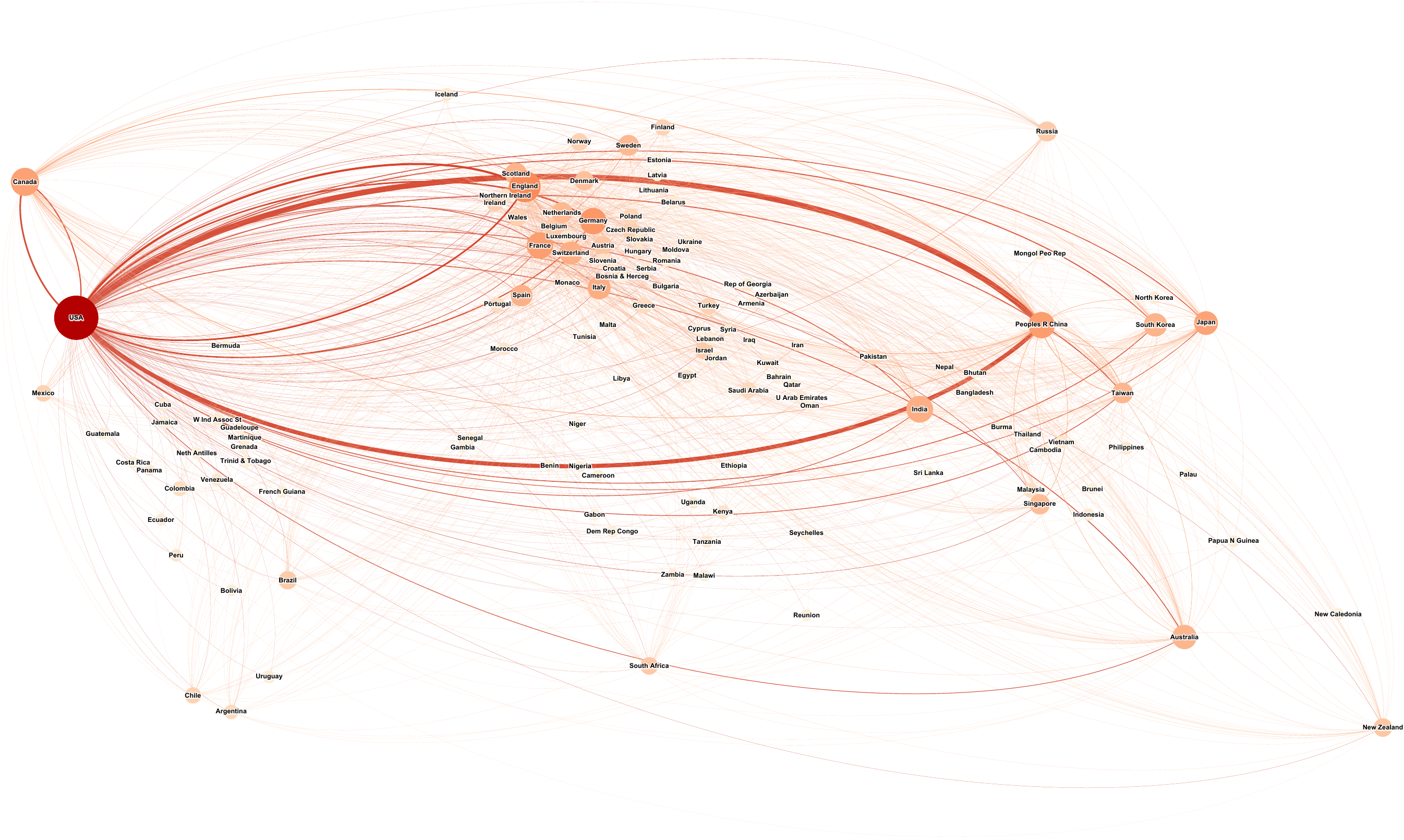}
		\label{fig:usa}
		\caption{Paths for super-movers from country of academic origin USA (high-resolution version online)}
\end{figure}

In Appendix~\ref{s:app}, we compare the mobility patterns of peripatetic researchers from different countries of academic origin by individually plotting the international paths of the super-movers from China, England, Germany, Japan, and the USA. The sizes of the nodes in Figures~\ref{fig:china}(6)--~\ref{fig:usa}(10) are proportional to their in-degrees, which equal the number of moves to that country. The out-degrees of the nodes, represented by darker shades, equal the number of moves out of a country. The direction of the curved edges is clockwise.

\begin{figure}[hbt!]
\centering
		\includegraphics[height=0.36\textheight]{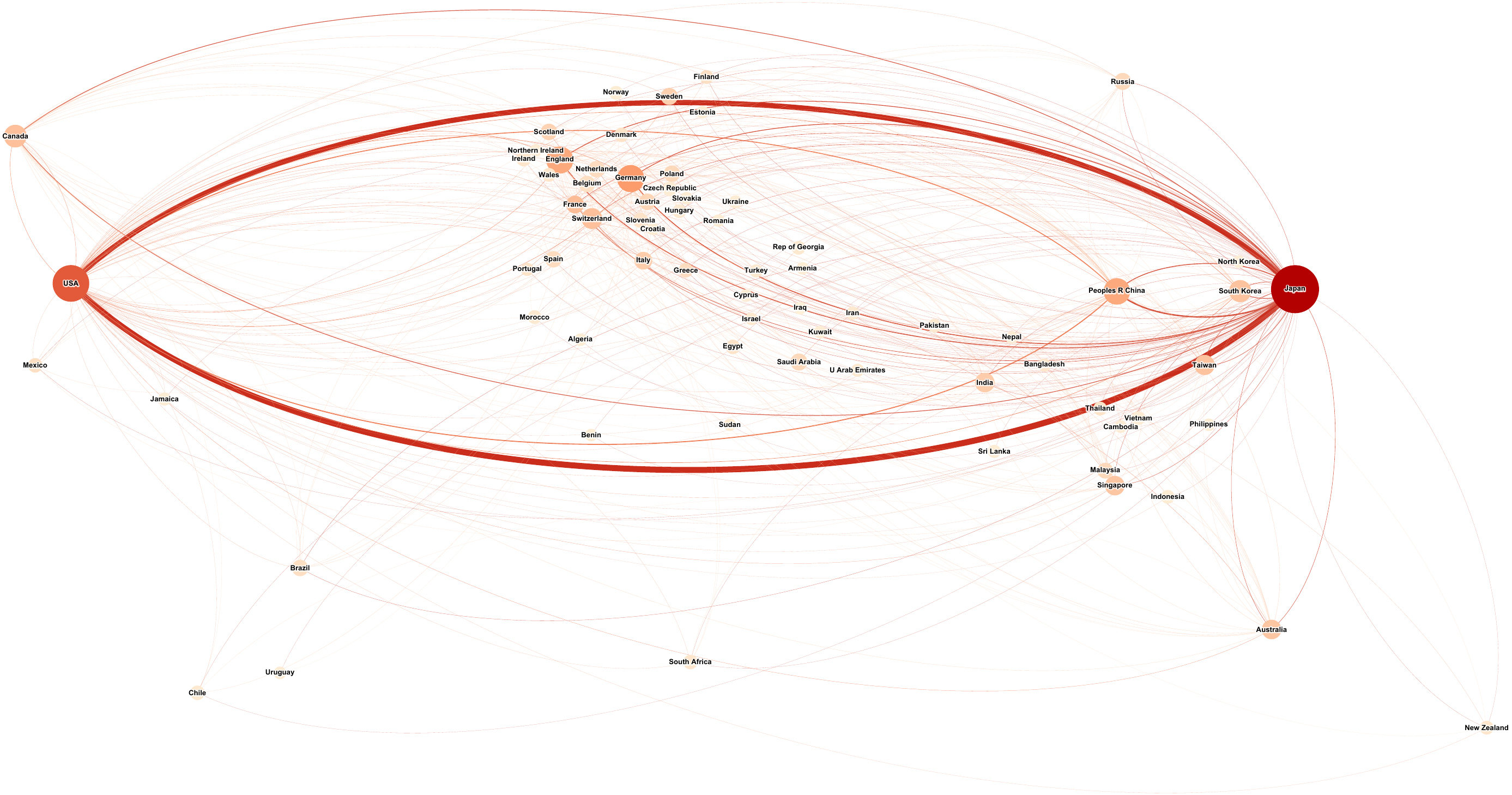}
		\label{fig:japan}
		\caption{Paths for super-movers from country of academic origin Japan (high-resolution version online)}
\end{figure}

Looking at Figures~\ref{fig:china}(6)--~\ref{fig:japan}(9), we can see that the USA was the country that the super-movers from China, England, Germany, and Japan moved in and out of second-most frequently. We list the countries that the super-movers moved in and out of third- to fifth-most frequently in decreasing order of frequency. The super-movers from China also frequently moved in and out of Taiwan, Canada, and Singapore. The super-movers from England were especially likely to move in and out of China, Switzerland, and Germany. We also observe that the super-movers from Germany frequently moved in and out of Switzerland, England, and France. The super-movers with an academic origin in Japan had frequent moves to and from China, South Korea, and Australia. Figure~\ref{fig:usa}(10) shows the four countries that the super-movers from the USA most frequently moved in and out of to be China, England, Germany, and South Korea. 

We use betweenness centrality (which measures how often a specific node appears in the shortest path between two other nodes of the network \cite{freeman1978centrality}) to rank the countries in a network of all super-movers' paths. Countries with the highest betweenness centrality were the USA, England, France, Germany, and Australia respectively. It should be noted that China was ranked 18 which seems to suggest that while China is an important node in the global system of science, it has limited influence as a connector in the paths of highly mobile scholars.

\begin{figure}
	\centering
	\subfloat[Super-movers by number of countries]{
		\includegraphics[height=0.3\textwidth]{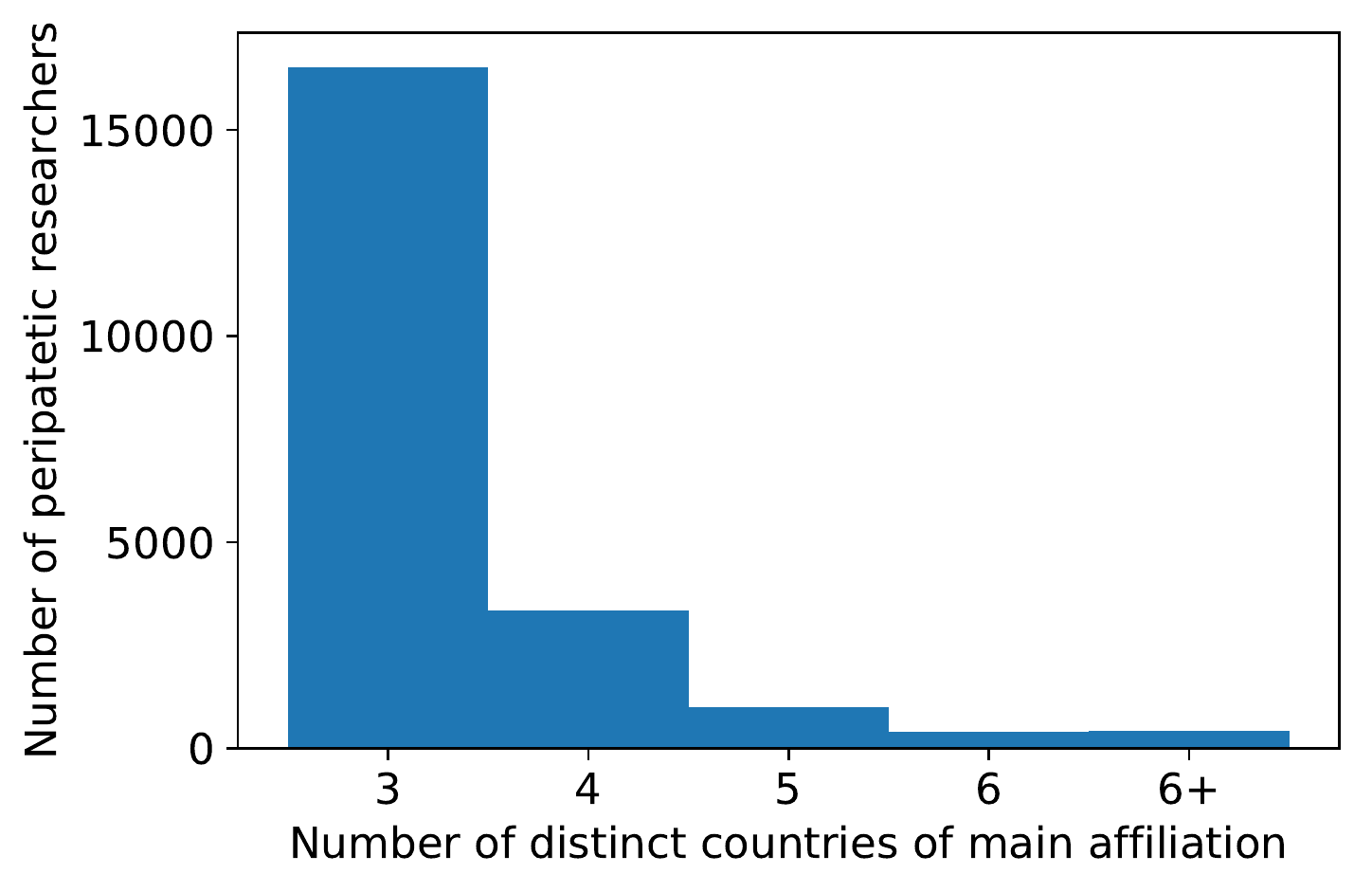}
		\label{subfig:affiliation_countries}
	}
	\hfill
	\subfloat[Super-movers by academic age]{
		\includegraphics[height=0.31\textwidth]{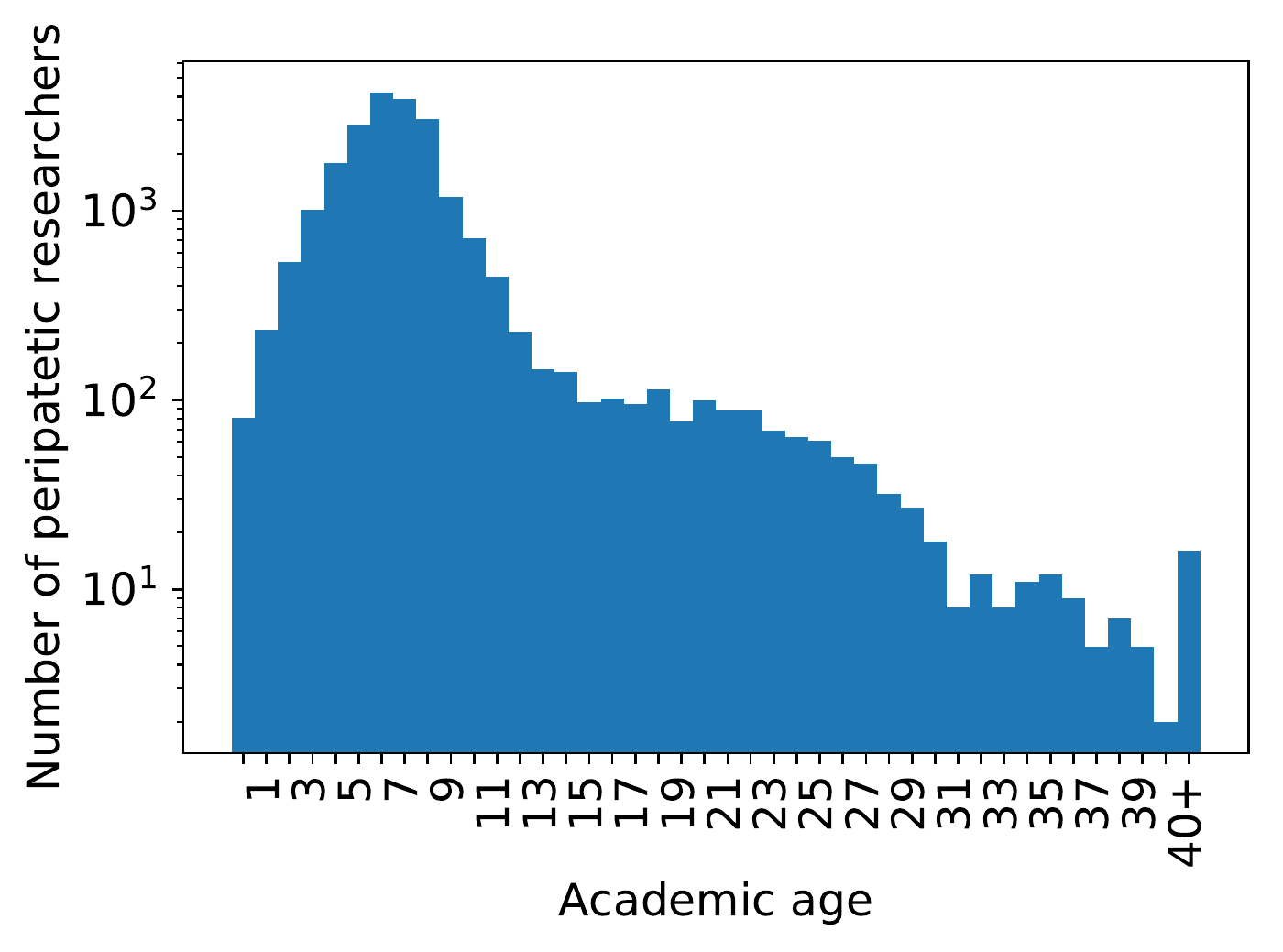}
		\label{subfig:academic}
	}
	\caption{Number of super-movers by their number of countries(a) and academic age(b)}
	\label{fig:log}	
\end{figure}

The data show that most of the peripatetic researchers had three countries of main affiliation in their international mobility paths. However, some have been affiliated with even more countries as shown in Subfigure~\ref{subfig:affiliation_countries}. 

We define the real-valued variable \textit{academic age} as the duration in years between the publication dates of the earliest and the latest authorship records of an individual scholar. Subfigure~\ref{subfig:academic} shows the academic ages of the peripatetic researchers. Note that the y-axis in Subfigure~\ref{subfig:academic} is on a logarithmic scale. The mode of academic age is in $[6,7]$ (similar to the findings of \cite{vaccario2018reproducing} for mobile scholars), but academic ages up to $10$ are also very frequent. Two major reductions in frequency are observed in the academic age brackets $10-11$ and $30-31$, the latter being possibly attributable to retirement.

The results displayed in Subfigure~\ref{subfig:academic} therefore suggest that the levels of experience among the peripatetic researchers in our data varies substantially. 
Based on the peaks and valleys of academic age observed in Subfigure~\ref{subfig:academic}, we continue our investigation by categorizing the authorship data into three brackets based on academic age: $[0,7)$ for early-career, $[7,14)$ intermediate-level, and $[14,+ \infty)$ senior super-movers. We observe in our data that most of the super-movers (49\%) belong to the early-career group. A slightly smaller but still substantial fraction (44\%) of the super-movers belong to the intermediate group, while a much smaller share (6\%) of the super-movers belong to the senior group. Given these observations, we provide in Subsection~\ref{ss:individual} several individual-level statistics with respect to academic age and country of academic origin, as the two main factors for disaggregating the authorship records of peripatetic researchers. 

\subsection{Destinations and return migration}\label{ss:individual}

In this subsection, we take the heterogeneity in levels of experience into consideration and accordingly provide an individual-level analysis of destinations and return migration. 

\begin{figure}
\centering
\subfloat[Destinations for academic age $[0,7)$]{
\includegraphics[width=0.48\textwidth]{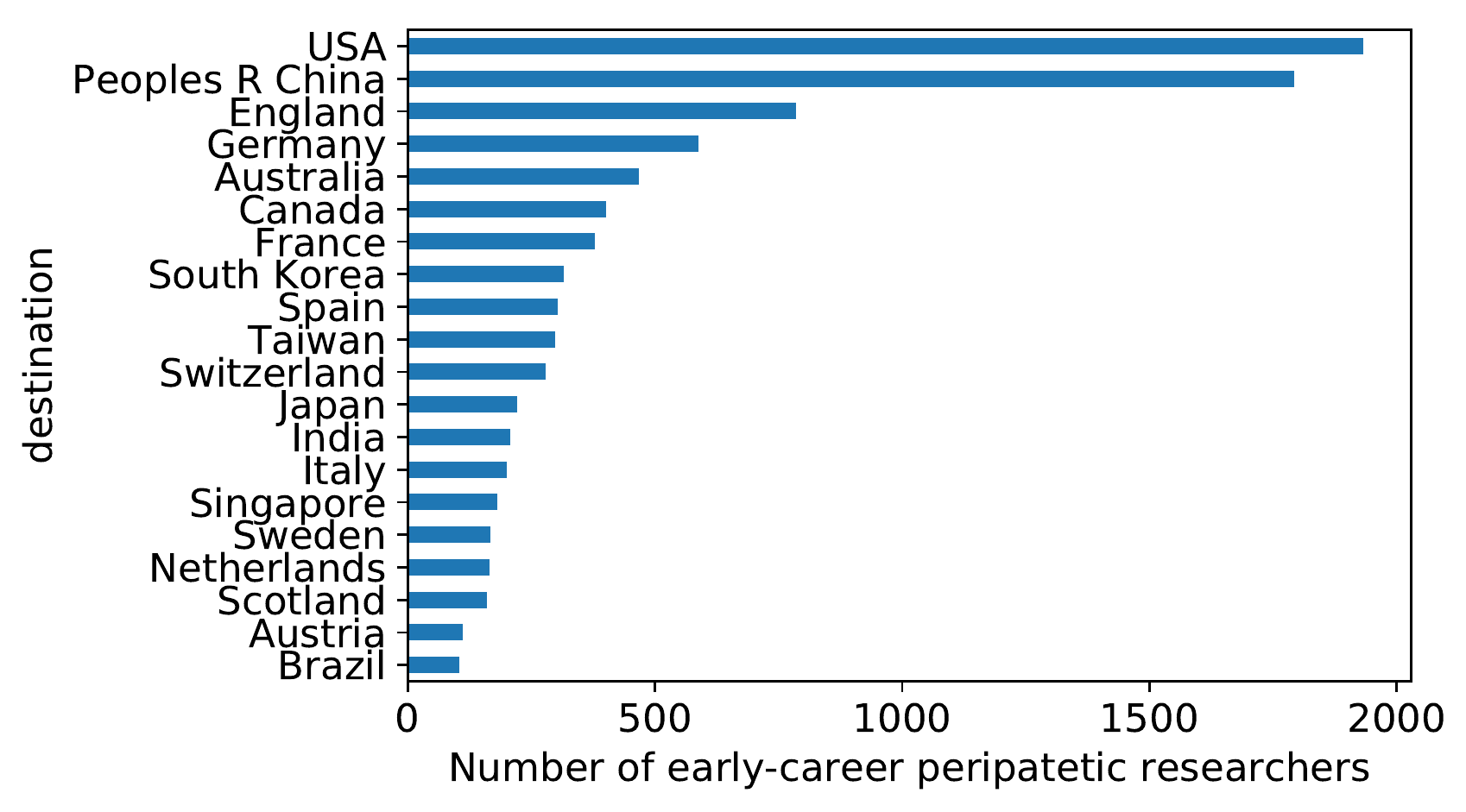}
\label{subfig:destination0}
}
\hfill
\subfloat[Destinations for academic age $[7,14)$]{
\includegraphics[width=0.48\textwidth]{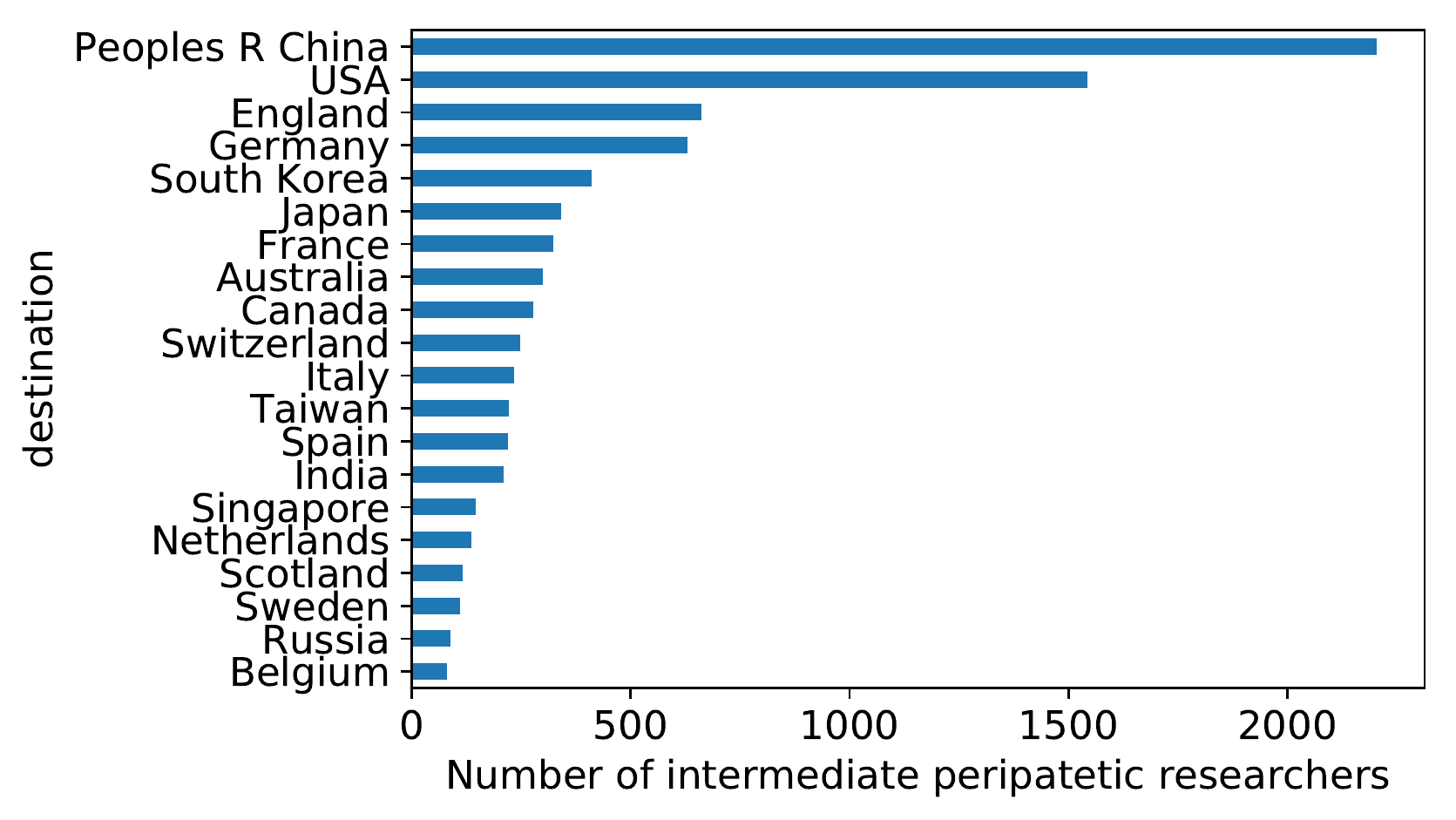}
\label{subfig:destination7}
}

\subfloat[Destinations for academic age $[14,+\infty)$]{
\includegraphics[width=0.48\textwidth]{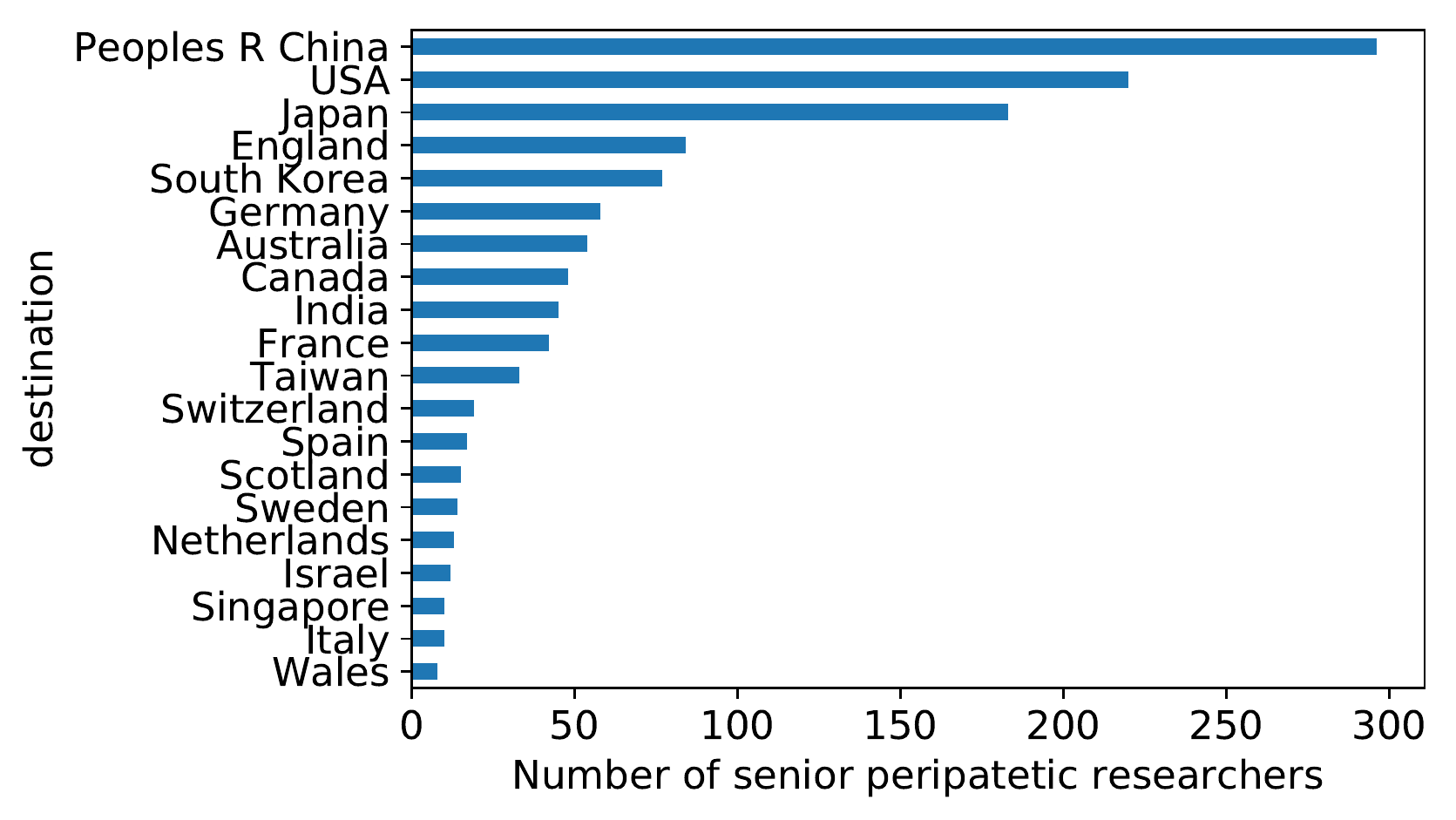}
\label{subfig:destination14}
}
\caption{Most common countries of academic destination among super-movers by academic age within the ranges of (a) $[0,7)$, (b) $[7,14)$, and (c) $[14,+\infty)$} \label{fig:destination_by_age}
\end{figure}

Figure~\ref{fig:destination_by_age} shows the most common countries of academic destination for the three academic age brackets. Subfigure~\ref{subfig:destination0} shows that the USA was the most common country of destination for the early-career super-movers (academic age up to seven) by a small margin, while Subfigures~\ref{subfig:destination7} and \ref{subfig:destination14} show that China was, by a large margin, the most common country of destination for both the intermediate super-movers (academic age between seven and 14) and the senior super-movers (academic age above 14). The third- and the fourth-most common countries of academic destination for both the early-career and the intermediate super-movers were, respectively, England and Germany; while the third- and the fourth-most common countries of academic destination for the senior super-movers were, respectively, Japan and England. As Figure~\ref{fig:destination_by_age} shows, certain countries, such as South Korea and Japan, were popular destinations for the more senior super-movers; while other countries, such as Australia, Canada, and Spain, were especially common as destinations for more early-career super-movers. 

We use the international mobility paths of each peripatetic researcher to check whether they had returned to their country of academic origin. For each country of academic origin $X$, we quantify \textit{return migration} as a simple fraction of the number of super-movers who had country $X$ in both their earliest and their latest publications to the total number of super-movers from country of academic origin $X$. 

Note that return migration aggregated for all countries equals $0.28$, $0.49$, and $0.45$ for the age brackets $[0, 7)$, $[7, 14)$, and $[14, +\infty)$, respectively; which seems to suggest that the intermediate and the senior super-movers were more inclined to return to the country where they first published than the early-career super-movers, who may have been visiting several countries as part of their professional development. 

\begin{figure}[hbt!]
\centering
\subfloat[Return migration for academic age up to 7 (early-career)]{
\frame{\includegraphics[width=0.97\textwidth]{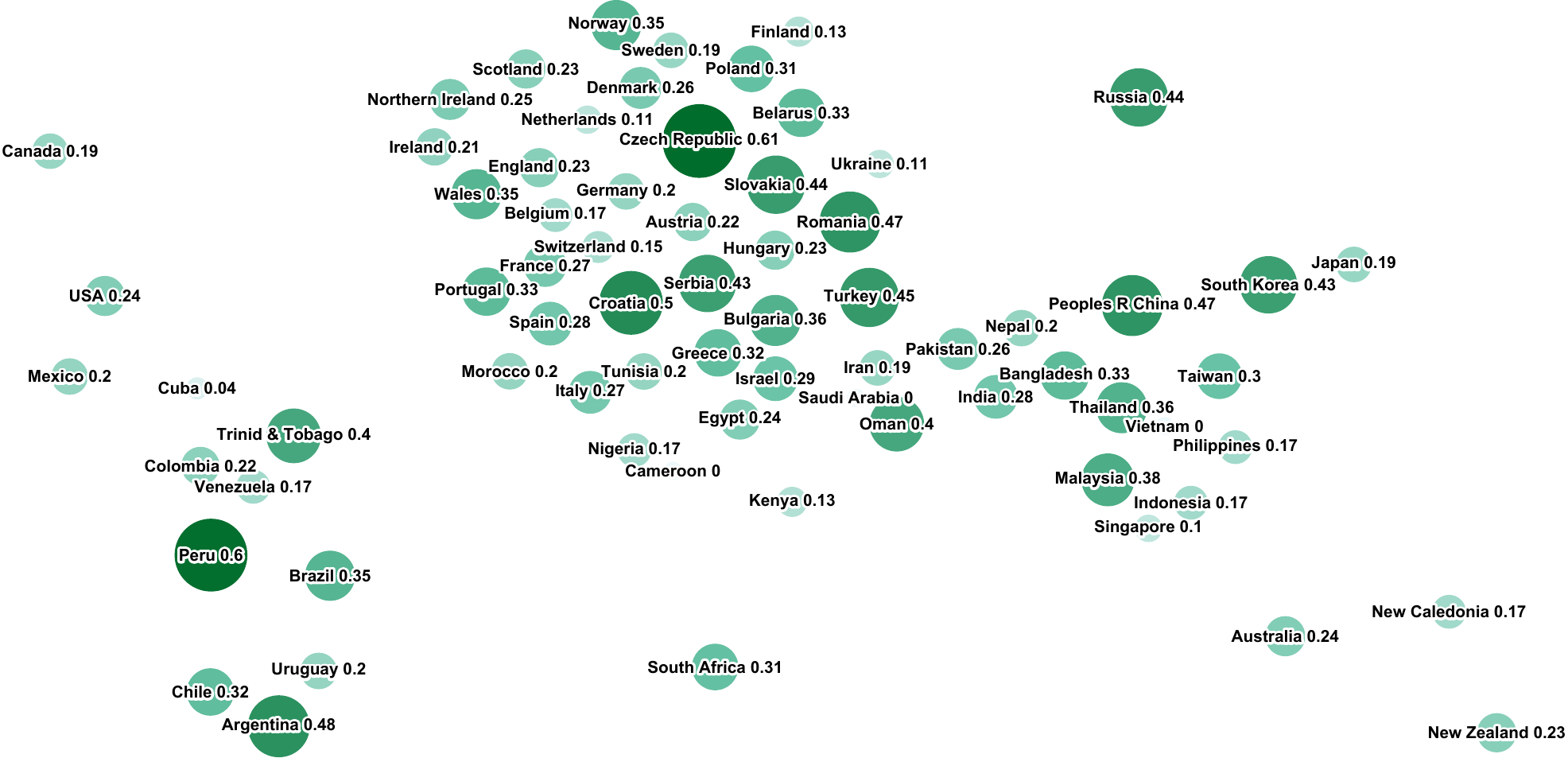}
\label{subfig:return0}}
}
\hfill
\subfloat[Return migration for academic age between 7 and 14 (intermediate)]{
\frame{\includegraphics[width=0.97\textwidth]{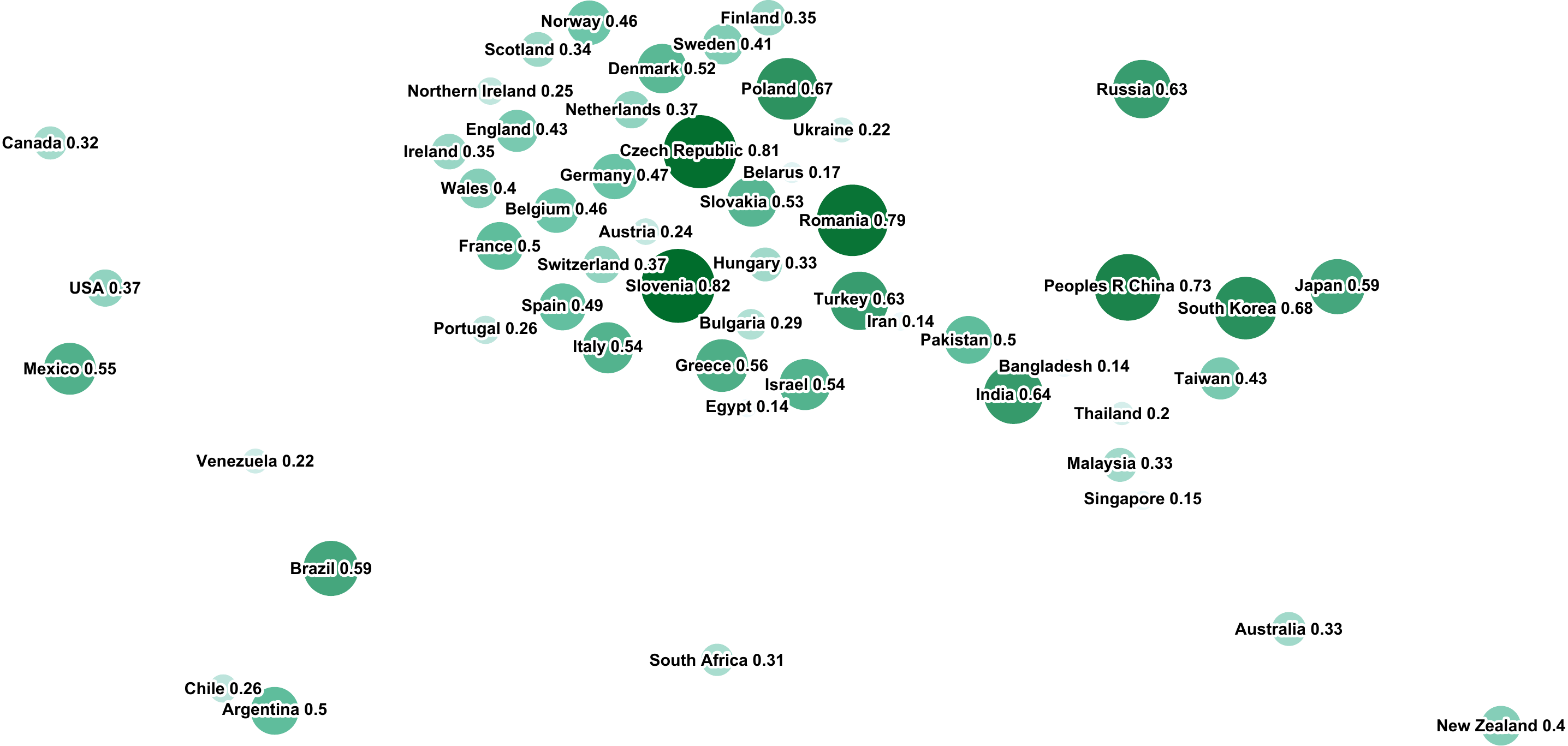}
\label{subfig:return7}}
}

\caption{Fraction of return migration by country of academic origin and academic age bracket: (a) early-career and (b) intermediate (high-resolution version online)} \label{fig:return}
\end{figure}

Figure~\ref{fig:return} shows the fractions of return migration among the early-career and the intermediate super-movers for different countries of academic origin, with larger circles and darker shades representing a larger value of return migration. In this analysis, we have omitted all of the countries for which the number of super-movers within the respective academic age bracket was less than five. The exact numeric values for return migration are provided next to the names of the countries in Figure~\ref{fig:return}. Subfigure~\ref{subfig:return0} shows that no early-career super-mover from Cameroon, Saudi Arabia, or Vietnam had returned to his/her country of academic origin. 

Looking at Figure~\ref{fig:return}, we can see that for both the early-career and the intermediate super-movers, levels of return migration were relatively low in Iran, Singapore, Ukraine, and Venezuela; and were medium-low in countries such as Australia, Austria, Canada, Egypt, Finland, Hungary, Ireland, the Netherlands, Scotland, Switzerland, and the USA. Levels of return migration were medium in Denmark, England, France, Germany, Israel, Italy, Norway, Pakistan, Spain, and Taiwan. The countries with medium-high levels of return migration were Brazil, Greece, Poland, and Slovakia; while the countries with high levels of return migration were the Czech Republic, China, Romania, Russia, South Korea, and Turkey.

Comparing to the average values, we observe that for seven countries -- Bangladesh, Bulgaria, Belarus, Malaysia, Portugal, Thailand, and Wales -- levels of return migration were relatively high among the early-career researchers, but were relatively low among the intermediate super-movers. For Japan and Mexico, the opposite pattern is observed: levels of return migration were relatively low among the early-career and relatively high among the intermediate super-movers.

Our results can be combined with country-specific fractions of researchers working abroad. For instance, the early-career return migration value for India becomes particularly informative when combined with the existing knowledge that 40\% of India-born researchers below 37 years of age have been employed outside of India \cite{toma_internationalization_2018}.

\section{Limitations and advantages of using bibliometric data}\label{ss:limit}

While using new data sources for studying mobility of researchers seems promising \cite{vaccario2018reproducing, robinson-garcia_many_2019}, we would like to remind the reader that there are some well-known challenges associated with relying on bibliometric data that our study does not resolve \cite{appelt_which_2015, robinson-garcia_many_2019}. For example, the time it takes to conduct research and publish papers is an essential factor that should not be neglected. This time lag prevents us from observing mobility events at the exact point in time when they occurred. Our analysis and results were solely based on main affiliation (as opposed to considering multiple affiliations \cite{robinson-garcia_many_2019}) which does not always represent the actual geographic location of a researcher. Furthermore, we are unable to observe and track international mobility not represented in Web of Science publication data, which are known to be biased toward certain languages, and to underrepresent certain countries \cite{sugimoto2018measuring}. We should, therefore, stress that our analysis is based on the underestimate of mobility that is currently achievable within the limits of bibliometric data. Future methodological research on scientometrics and complementary data sources could, perhaps, address these issues. 

An important technical limitation of the data is related to author disambiguation. We used the author IDs that Web of Science offers and performed sensitivity analyses based on a set of 7,000 manually disambiguated authorship records. Although the substantive results of our paper did not change, we observed that author disambiguation represents an important challenge to the validity of the dataset. Thus, we believe that new and innovative methods of algorithmic author name disambiguation are needed to improve the overall quality of bibliometric data. 

Despite these limitations, it is also important to point out that by using bibliometric data, we have been able to conduct an analysis of mobility that is more cross-disciplinary, longitudinal, globally scaled, and contemporary \cite{czaika_globalisation_2018, robinson-garcia_many_2019, alburez2019} than studies that use classic data sources, such as registries, surveys, and censuses. Most of the studies that use bibliometric data, including this one, have the advantage of being easily replicable at different levels of analysis. For instance, our framework can be adopted for investigating scientific mobility at a national or a regional level, across scientific fields, and across research institutes and universities. 

\section{Discussion and future directions}\label{s:future}

In this paper, we provided for the first time (to the best of our knowledge) a snapshot of the characteristics of highly mobile scholars (super-movers), whom we define as those researchers who have had main affiliations in at least three distinct countries. Our goal was to identify the common features that distinguish a highly selected group of researchers who are in some ways still ``outliers,'' but who also serve as the oil that lubricates the global system of brain circulation in science. We have witnessed the emergence of a system that includes the USA and China as two large hubs, and England and Germany as two smaller hubs for highly mobile scholars. It is important to note that, despite the bias toward English-speaking journals in the Web of Science, China was the top country of destination among the super-movers in the dataset. This may be an indication of the progress China has made in reaching its goal of becoming a major science powerhouse. 

Demographic perspectives have seldom been considered in previous bibliometric research. With this study, we add a demographic flavor to the science of science literature by also accounting for the age patterns of mobility, and by considering metrics (like return migration) that are common in demographic studies, but are not typically considered in the science of science literature. Accounting for the age distribution of scholars is essential to avoid obtaining spurious results that are affected by compositional changes in the underlying population. Among the aims of this article is to foster the development of bridges between demographic and scientometric research. We expect that fruitful interdisciplinary collaborations will emerge and evolve in the future. 

This study represents the initial step in a continuing effort to construct a longitudinal dataset on the mobility of researchers and on the geographical trajectories of their career paths. The data on these trajectories could be used to generate migration estimates that facilitate the investigation of phenomena such as brain drain and brain circulation. In addition, by leveraging the networked structure of the data, it may be possible to test and advance network theories of international migration \cite{massey_theories_1993}. 

We developed a panel of fine-grained data on the geographic trajectories of highly mobile scholars, which are intended to stimulate further research at the intersection of migration research and scientometrics. On the one hand, the data can be used to address the question of to what extent countries' shares of brain circulation depend on factors such as the diversity of their science system and national-level measures of research quality. On the other hand, the data can be used to study the effects of international mobility on measures of research quality and impact for individuals, organizations, and countries. 

\section*{Acknowledgements}
The authors thank the anonymous referees for their comments, and Chowdhury Majedur Rahman for assistance with the data quality checks. SA and EZ designed and conducted the research and wrote the paper. JW contributed in extracting the data.


\bibliography{references}
\bibliographystyle{splncs04}

\begin{subappendices}
	\renewcommand{\thesection}{\Alph{section}}%

\section{Paths of the peripatetic researchers}\label{s:app}

\begin{figure}[hbt!]
	\centering
		\includegraphics[height=0.4\textheight]{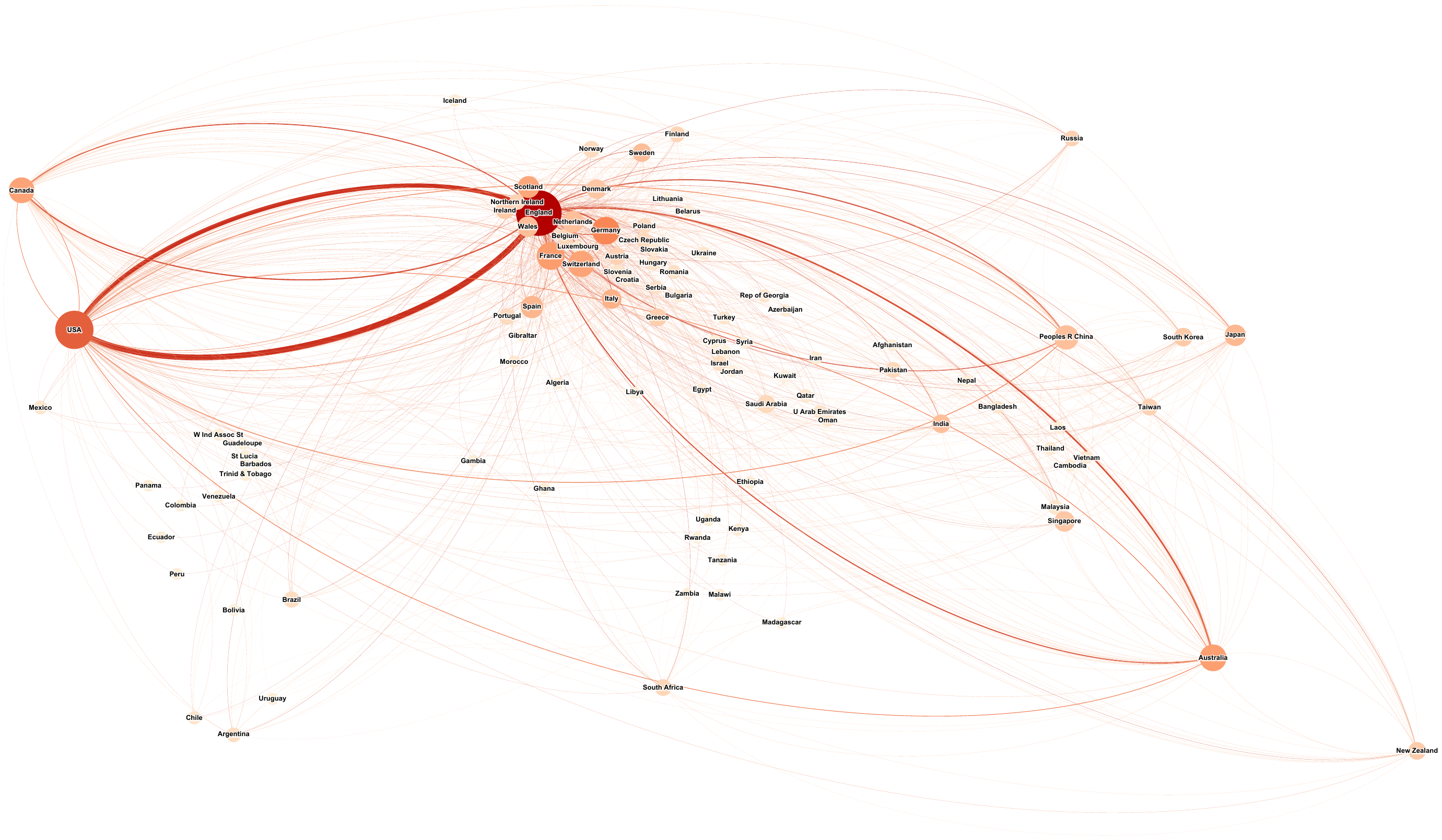}
		\label{fig:england}
		\caption{Paths for super-movers from country of academic origin England (high-resolution version online)}
\end{figure}

\begin{figure}[hbt!]
\centering
		\includegraphics[height=0.4\textheight]{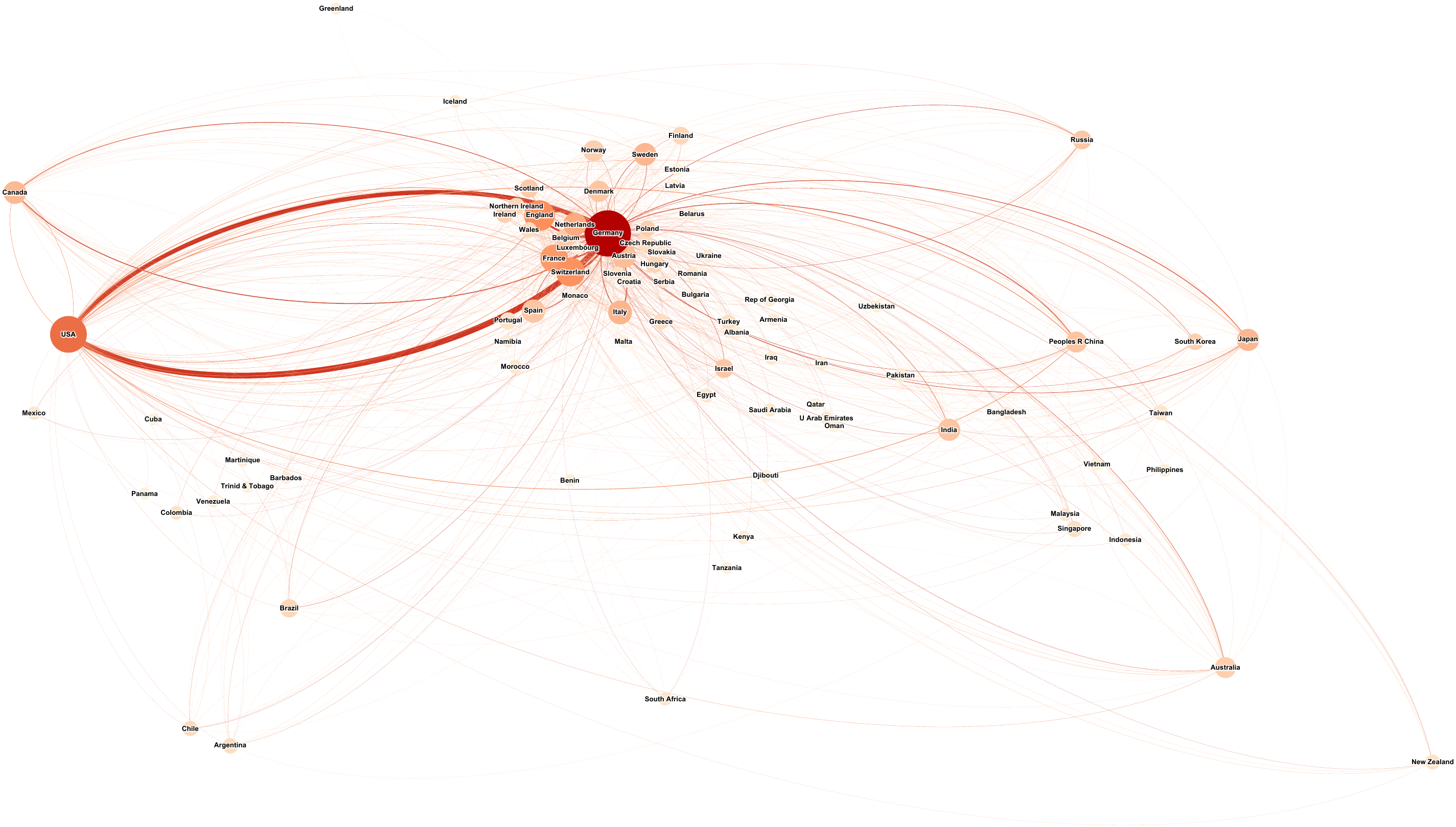}
		\label{fig:germany}
		\caption{Paths for super-movers from country of academic origin Germany (high-resolution version online)}
\end{figure}

\end{subappendices}

\end{document}